# Delayed elastic contributions to the viscoelastic response of foams

François A. Lavergne,[1,a)] 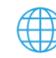 Peter Sollich,[2,3]  and Véronique Trappe[1]

**AFFILIATIONS**

[1] Department of Physics, University of Fribourg, Chemin du Musée 3, 1700 Fribourg, Switzerland
[2] Institute for Theoretical Physics, University of Göttingen, Friedrich-Hund-Platz 1, 37077 Göttingen, Germany
[3] Department of Mathematics, King's College London, Strand, London WC2R 2LS, United Kingdom

**Note:** This paper is part of the JCP Special Topic on Slow Dynamics.
[a)] Author to whom correspondence should be addressed: francois.lavergne@unifr.ch

**ABSTRACT**

We show that the slow viscoelastic response of a foam is that of a power-law fluid with a terminal relaxation. Investigations of the foam mechanics in creep and recovery tests reveal that the power-law contribution is fully reversible, indicative of a delayed elastic response. We demonstrate how this contribution fully accounts for the non-Maxwellian features observed in all tests, probing the linear mechanical response function. The associated power-law spectrum is consistent with soft glassy rheology of systems with mechanical noise temperatures just above the glass transition [Fielding *et al.*, J. Rheol. **44**, 323 (2000)] and originates from a combination of superdiffusive bubble dynamics and stress diffusion, as recently evidenced in simulations of coarsening foam [Hwang *et al.*, Nat. Mater. **15**, 1031 (2016)].

## I. INTRODUCTION

Liquid foams are dense packings of bubbles dispersed in a surfactant-rich continuous phase.[1] At large enough packing fractions, the foam elasticity is determined by the resistance of bubbles to be deformed, which is proportional to the surface tension and inversely proportional to the bubble radius.[1,2] However, unlike other over-jammed systems, foams are not only elastic. Indeed, due to differences in Laplace pressure between small and large bubbles, foams coarsen in time.[1,3] This process continuously perturbs the stress balance between contacting bubbles, eventually resulting in configurations that exceed the local yield conditions, which in turn triggers local bubble rearrangements.[4–6] This structural relaxation mechanism provides the means of relaxing macroscopically imposed stresses[4,7] such that foams typically display a very slow fluid-like response.[2,7,8] Assuming that bubble rearrangement events occur stochastically,[6] one would *a priori* expect stress relaxation to be exponential, a characteristic of "Maxwell" fluids.

However, the slow viscoelastic response of foams strongly differs from that of simple "Maxwell" fluids that are solely characterized by an instantaneous elastic response and a single relaxation time.[9] Indeed, while the foam responds to the application of a small constant stress by an instantaneous elastic deformation, an intermediate creep phase precedes the terminal relaxation.[2,8] Similarly, the frequency-dependent response to an oscillatory strain is characterized by a storage modulus that gradually decays with decreasing frequencies, before exhibiting the hallmarks of viscous relaxation.[2,7,10,11]

In this work, we show that these non-Maxwellian features relate to a time-dependent creep strain that is reversible. This delayed elastic contribution displays a power-law behavior, which is directly probed in strain recovery experiments following the release of a constant applied stress. We find that accounting for delayed elastic contributions in addition to Maxwellian behavior fully describes the linear mechanical response functions of foam at long times and equivalently at low frequencies. Our work thus identifies foam as a power fluid with a final relaxation, reminiscent of systems near the jamming transition. Within the class of over-jammed systems, this feature denotes foams as unique: they are over-jammed, yet do relax due to coarsening-induced structural relaxation events, which govern both delayed elastic storage and dissipation.

## II. EXPERIMENTAL

Our experimental system is a Gillette shaving cream with coarsening characteristics similar to those of Gillette foams used in numerous previous studies.[4–8,10] It comprises propellant gases isobutane and propane, dispersed at a volume fraction of $\phi = 91.5\% \pm 0.3\%$ in an aqueous mixture of long-chain fatty acids and sodium lauryl sulfate. All rheological tests are performed at 25 °C starting at $t_w$ = 5000 s, with $t_w$ defining the time elapsed since the foam production. By that time, the average bubble radius is $52 \pm 5$ $\mu$m, and the foam has evolved to the dynamical scaling regime of coarsening,[1,3,11] as shown in Fig. S1 of the supplementary material.

Rheological experiments are performed by using a stress-controlled rheometer from Anton Paar (MCR series), equipped with a wide-gap Couette geometry (gap size 5.02 mm), which ensures that the mechanics probed is that of a bulk foam. Both cup and cylinder of the Couette cell are sanded to avoid wall slip, and the cell is sealed using a solvent trap to minimize drying and excessive gas exchange with the environment. The foam is loaded right after production through a hole at the bottom of the cup with the cylinder locked in measuring position, as described in Ref. 7. To minimize drift effects due to the air-bearing connection of the cylinder to the motor, we set the measuring position to the angle at which the residual stress observed after releasing a constant stress applied on a viscous fluid (glycerol) is minimal (<$10^{-4}$ Pa).[12] Such precaution becomes particularly important in recovery experiments, where residual torques may affect the signals when the shear rates approach the resolution limit.[12]

The following rheological tests are performed within the linear range of viscoelasticity as demonstrated in Fig. S4 of the supplementary material: (i) stress relaxation tests measuring the temporal evolution of the stress $\sigma(t)$ after application of a step strain of $\gamma_0 = 0.1\%$, (ii) oscillatory strain experiments measuring the frequency dependence of the storage $G'(\omega)$ and loss modulus $G''(\omega)$ with $\gamma_0 = 0.1\%$, and (iii) creep and recovery tests measuring the temporal evolution of the strain $\gamma(t)$ after applying a step stress of $\sigma_0 = 0.2$ Pa for a time of $t_c = 1000$ s, and subsequently resetting the stress to zero. This procedure guarantees quasi-stationary creep conditions as the bubble size does not change by more than 8% during stress application, as shown in Fig. S1(b) of the supplementary material. Since the oscillatory stain experiments typically last 5000 s, we correct the values of the moduli by taking into account the drop of the foam elasticity during the test, as detailed in Figs. S2 and S3 of the supplementary material. All experiments are repeated at least six times following the loading procedure described above, where we report the averaged data in the figures hereafter and use the standard deviation as a measure of error.

## III. RESULTS AND DISCUSSION

The results of all three tests clearly indicate that the foam is not simply characterized by an elastic modulus and a single relaxation process, as this is the case for Maxwell fluids.[9] Instead, stresses relax following a more complex pathway. This can be appreciated in Fig. 1(a), where we report the time dependence of the relaxation modulus $G(t) = \sigma(t)/\gamma_0$ obtained in stress relaxation tests. Estimating the foam viscosity as $\eta_R = \int_0^{+\infty} G(t)dt \approx 5 \times 10^4$ Pa. s and the short-time or equivalently high-frequency modulus as $G_\infty \approx 190$ Pa, the expected Maxwellian response would be $G(t) = G_\infty e^{-t/\tau_R}$, with a relaxation time of $\tau_R = \eta_R/G_\infty \approx 260$ s.[9] This description, depicted as a dashed line in Fig. 1(a), is, however, clearly inconsistent with our experimental findings. Deviations from Maxwellian characteristics are also apparent in the frequency dependence of $G'(\omega)$ and $G''(\omega)$, as shown in Fig. 1(b). Indeed, while the mechanical response of a Maxwell fluid is characterized by a sharp crossover between the high-frequency regime, where $G'(\omega)$ is independent of $\omega$ and $G''(\omega)$ decreases with increasing frequency, and the low-frequency range where both moduli increases, the foam displays a markedly different behavior. In the high frequency range, the loss modulus increases as a square root of frequency, a feature that has been attributed to nonaffine motion along random slip planes.[7,13] Moreover, deviations from the Maxwell behavior are also observed in the lower frequency range. Indeed, the storage modulus gradually decays upon decreasing the frequency, instead of being almost frequency independent before exhibiting the hallmarks of relaxation. While these observations are consistent with results of previous investigations,[2,7,8,10,11] the description of the low-frequency characteristics of $G'(\omega)$ and $G''(\omega)$, or equivalently the long-time behavior of $G(t)$, remains unclear.

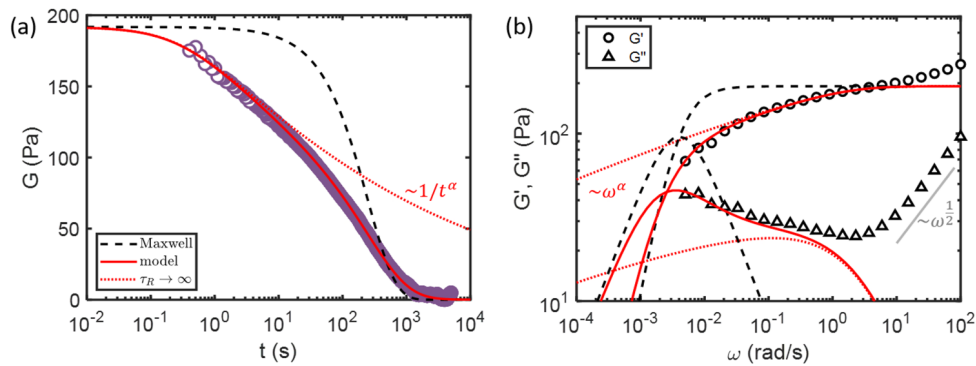

**FIG. 1.** (a) Time dependence of the relaxation modulus $G(t)$ with data averaged over seven runs and error bars comparable to the symbol size. (b) Frequency dependence of the storage $G'(\omega)$ and loss modulus $G''(\omega)$ with data averaged over six runs and error bars comparable to the symbol size. Dashed-black lines indicate the behavior of a Maxwellian fluid with magnitudes of $G_\infty$ and $\eta_R$ corresponding to those of the foam. Red solid lines denote the behavior expected from the description of the creep data, using Eqs. (5) and (6); red dotted lines show the behavior expected for $\tau_R = \eta_R/G_\infty \to \infty$. The gray solid line depicts an increase of the moduli as $\sqrt{\omega}$ (gray solid line), as expected for dense particle packings in the range of high frequencies.[7,13]

New insight is gained by examining the results of the creep and recovery tests. Figure 2(a) displays the time dependence of the strain during creep and recovery, normalized by the stress $\sigma_0$ applied during the creep test, $J(t) = \gamma(t)/\sigma_0$. Formally, $J(t)$ is the compliance for data acquired in creep tests, while for recovery, the normalization with $\sigma_0$ is a convenient way to compare creep and recovery responses within the same framework. For the creep experiment performed by applying a constant stress, the foam elasticity appears as an instantaneous rise of $J = 1/G_\infty$, with oscillations due to the tool inertia.[14] This is followed by an increase of $J(t)$ that reflects the foam relaxation processes. As expected, a compliance describing the behavior of Maxwell fluids $J(t) = 1/G_\infty + t/\eta_R$ fails to describe the data.[9] Reporting the derivative of $J(t)$ as a function of time reveals that this discrepancy corresponds to a gradual decrease of the compliance rate, which reaches the constant value $1/\eta_R$ only at long times, as shown in Fig. 2(b).

The nature of this decrease is uncovered upon releasing the stress in recovery tests. Indeed, beyond the elastic recoil of $1/G_\infty$ expected for a Maxwell fluid, we find that $J(t)$ further decreases in time, as shown in Fig. 2(a). This delayed recovery is characterized by a power-law recovery rate, which matches the decay of the creep rate, as shown in Fig. 2(b). This suggests that the initial creep is also best described by a power law and more importantly that it is due to a time-dependent, yet recoverable contribution, thereby indicating a delayed elastic contribution to creep.[15–17]

To describe our data, we thus write the creep compliance as a sum of three terms

$$J(t) = \frac{1}{G_\infty} + \frac{m}{G_\infty} \frac{(1 + t/\tau_0)^\alpha - 1}{\alpha} + \frac{t}{\eta_R}, \quad (1)$$

where the delayed elastic contribution is described by the Jeffreys–Lomnitz law of amplitude $m$ and exponent $0 < \alpha < 1$.[18,19] Note that for $t \gg \tau_0$, the Jeffreys–Lomnitz law reduces to a simple power law of $J(t) \sim t^\alpha$. As shown in Figs. 2(a) and 2(b), Eq. (1) fits our data remarkably well, yielding $\eta_R = 4.5(\pm 0.6) \times 10^4$ Pa·s and $G_\infty = 192 \pm 7$ Pa, in good agreement with the values inferred from the analysis of the stress relaxation test. For the parameters of the central term, we find $\tau_0 = 0.3 \pm 0.1$ s, $m = 0.09 \pm 0.01$, and a rather small creep exponent $\alpha = 0.17 \pm 0.03$. For recovery times small compared to the creep test duration $t_c = 1000$ s, we expect that recovery is well approximated by a simple reversal of the elastic contributions,

$$J(t) = J(t_c) - \frac{1}{G_\infty} - \frac{m}{G_\infty} \frac{(1 + t/\tau_0)^\alpha - 1}{\alpha}, \quad (2)$$

with $J(t_c)$ being the compliance reached at the end of the creep experiment and $t$ being the experimental timescale of the recovery test.[9] The good agreement between this description and the experimental data shown in Figs. 2(a) and 2(b) corroborates that short-time creep is caused by a delayed elastic storage process, which is reversible.

The power-law behavior suggests that this contribution is due to a process with a broad distribution of timescales, which appears naturally when describing the central term of Eq. (1) as a series of Kelvin–Voigt elements,

$$J(t) = \frac{1}{G_\infty} + \frac{m}{G_\infty} \int_0^{+\infty} \rho(\tau)(1 - e^{-t/\tau}) d\tau + \frac{t}{\eta_R}, \quad (3)$$

where $\rho(\tau)$ denotes the retardation spectrum.[9,19,20] This spectrum can be obtained as $\rho(\tau) = \frac{G_\infty}{m\tau} \mathcal{L}^{-1}[\dot{J} - 1/\eta_R]$, where $\mathcal{L}^{-1}$ denotes the inverse Laplace transform from the $t$-domain back to the conjugate variable $\lambda = 1/\tau$. For Eq. (1), such Laplace inversion has the following analytical solution:[19]

$$\rho(\tau) = \frac{e^{-\tau_0/\tau}}{\tau_0 \Gamma(1-\alpha)} \left(\frac{\tau_0}{\tau}\right)^{1-\alpha}, \quad (4)$$

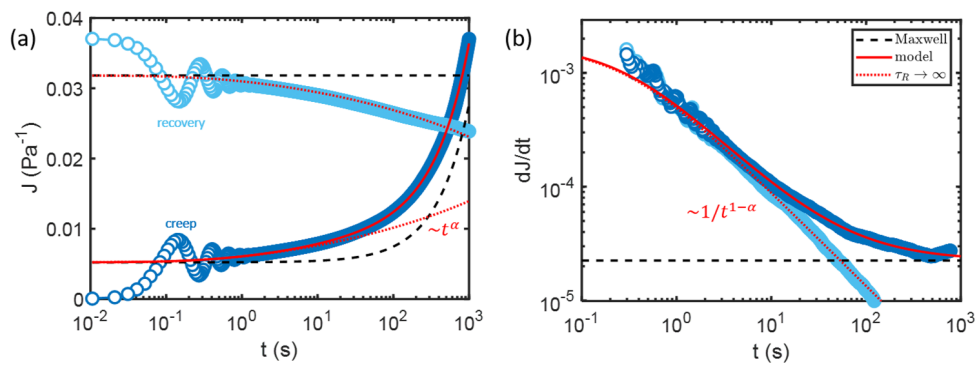

**FIG. 2.** (a) Temporal evolution of the strain during creep and recovery normalized by the stress applied during creep. (b) Same data as in (a) reported as creep and recovery rates. For the recovery rates, we limit the range of data shown to the shear rates that can be reliably measured with our rheometer. Dark and light blue symbols denote the experimental data, obtained in creep and recovery tests, respectively; the data are averages over eight runs with error bars comparable to the symbol size. Dashed-black lines indicate the behavior of a Maxwell fluid with the characteristics of the foam. Red solid lines denote the fit of the creep data to Eq. (1), correcting for tool inertia as described in the supplementary material. The resulting fit parameters are those used to describe the results of all other tests shown in this paper. The bottom red-dotted line in (a) corresponds to the creep response of the material with $\tau_R = \eta_R/G_\infty \to \infty$ in Eq. (1). The top red-dotted line in (a) corresponds to recovery according to Eq. (2) and the red-dotted line in (b) is the corresponding derivative.

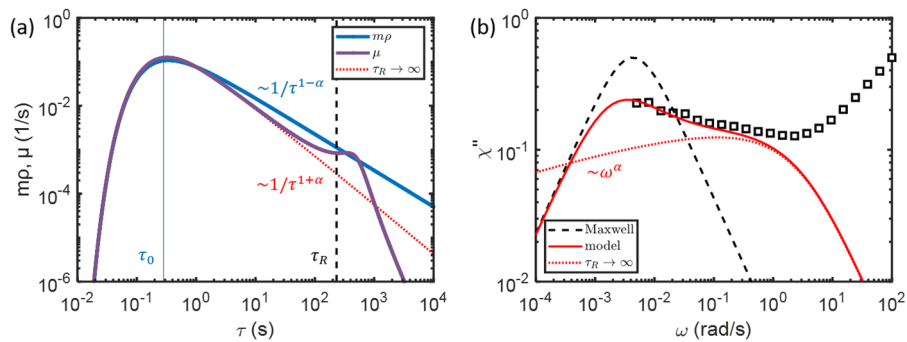

**FIG. 3.** (a) Retardation spectrum $\rho$ calculated using Eq. (4) (blue line), and corresponding relaxation spectrum $\mu$ calculated using Eq. (6) (purple line). The vertical thin blue line indicates the short-time delayed elastic cutoff of $\tau_0 = 0.3$ s; the vertical black-dashed line indicates the Maxwellian relaxation time of $\tau_R \approx 230$ s. Note that the peak in the relaxation spectrum is at $\approx 500$ s, considerably above $\tau_R$. The red-dotted line is the relaxation spectrum expected for $\tau_R \to \infty$. (b) Dynamic susceptibility $\chi''(\omega) = G''(\omega)/G_\infty$ as a function of frequency. Symbols denote the experimental data corresponding to those shown in Fig. 1(b). Red-solid line shows the behavior expected from the description of the creep data. Maxwell relaxation (dashed-black line) leads to the appearance of a low-frequency relaxation peak on top of the delayed-elastic "wing" corresponding to $\tau_R \to \infty$ (red-dotted line).

displaying a broad power-law tail of $\rho(\tau) \sim 1/\tau^{1-\alpha}$, as shown in Fig. 3(a).

To fully validate our modeling, we examine whether the stress relaxation of our foam can be described within the same framework. Since $\rho \neq 0$, we need to assume that stress relaxation is governed by a distribution of relaxation times,[9,17,20,21]

$$G(t) = G_\infty \int_0^{+\infty} \mu(\tau) e^{-t/\tau} d\tau, \quad (5)$$

where the relaxation spectrum $\mu$ relates to $\rho$ as[9,20,21]

$$\mu(\tau) = \frac{m\rho(\tau)}{\left(1 - \frac{\tau}{\tau_R} + \tau \int_0^{+\infty} \frac{m\rho(u)}{\tau - u} du\right)^2 + (\pi \tau m \rho(\tau))^2}, \quad (6)$$

taking the integral over $u$ as a principal value.[21] With $\rho$, $m$, and $\tau_R = \eta_R/G_\infty$ known from the fit of the creep data, we compute $\mu$ reported in Fig. 3(a) and then calculate $G(t)$ using Eq. (5). The resulting $G(t)$ describes the experimental data remarkably well, as shown in Fig. 1(a). Equivalently, we can determine $G'(\omega)$ and $G''(\omega)$ as a function of frequency by again summing up the Maxwell responses over the relaxation spectrum $\mu$.[9] As shown in Fig. 1(b), the result accounts for the characteristics of the experimental data in the low frequency range of interest. Note that the result is also more accurate than that obtained by using the usual Fourier-based method to transform $G(t)$ into $G^*(\omega)$,[9] as shown in Fig. S4(d) of the supplementary material. The fact that we can use the sole description of the creep data by Eq. (1) to accurately recast the results obtained in stress relaxation and oscillatory shear tests unambiguously shows that the process leading to time-dependent reversible strain in recovery is at the origin of the slow non-Maxwellian relaxation of foam.

Compared to other conversion methods,[2,7,22] the approach used here has the advantage of being analytically solvable, providing details about the distribution of relaxation times $\mu$. As denoted in Eq. (6), this distribution solely depends on $\rho$, the parameter $m$, and the terminal relaxation time $\tau_R = \eta_R/G_\infty$. In the limit

$\tau_R \to \infty$, $\mu$ decays as $\mu(\tau) \sim 1/\tau^{1+\alpha}$, as shown in Fig. 3(a). This decay is consistent with the predictions of soft glassy rheology (SGR), assuming a mechanical noise temperature of $x = 1 + \alpha \approx 1.17$ just above the glass transition.[23–25] The corresponding relaxation modulus and low-frequency response are shown in Figs. 1(a) and 1(b), respectively. For our foam, however, $\tau_R$ is finite, resulting in a cutoff of the power law and the appearance of a peak in the spectrum at $\approx 500$ s. This timescale is considerably above the Maxwellian relaxation time obtained from the creep data ($\tau_R = 230$ s) and is consistent with the inverse crossover frequency inferred from the extrapolation to $G'(\omega) = G''(\omega)$ of the data shown in Fig. 2(b). Such a shift of the effective relaxation to lower frequencies is caused by the $\rho$-dependent integral in Eq. (6) and reflects that the delayed elastic contributions effectively delay terminal relaxation.

Note that the use of the Jeffreys–Lomnitz law for the central term in Eq. (1) is also motivated by the possibility of interpolating between the behavior of Maxwell fluids with $\alpha = 1$ and the logarithmic creep $\sim \ln(1 + t/\tau_0)$ with $\alpha \to 0$,[18,19] which is generally observed in glassy systems.[16,19,26] Conveniently, these two cases are also consistent with the predictions of SGR for the liquid phase ($x > 2$) and the glass phase ($x < 1$), respectively.[25] The Jeffreys–Lomnitz law, therefore, provides an empirical formula that allows us to obtain retardation spectra analytically for a range of soft materials where power-law or logarithmic creep is commonly observed.[16,26–29]

## IV. CONCLUSION

In conclusion, our experiments show that the rheology of foam under quasi-stationary conditions retains features of power-law fluids but eventually exhibits a final relaxation. The experimental approach adopted highlights the advantage of using creep and recovery tests where the power-law contribution is assessed unambiguously, in contrast to other rheological tests where the different elastic and dissipative contributions are not simply additives. Most importantly, the recovery tests show that power-law creep is related to a process that allows for delayed elastic recovery.[15–17] To account for this behavior, let us recall that the coarsening process in foam leads to intermittent restructuring events.[4–6] In creep experiments, a restructuring event leading to the relaxation of local stresses entails that the macroscopic stress will be carried by the rest of the system.[16] This results in an increase in strain that can only be recovered in time; this is because upon releasing the stress, a full recovery of the stress-bearing part of the system will be hindered by the zones that were previously relaxed and now resist the deformation imposed by the instantaneous elastic recoil. Further restructuring events will then be needed to allow for full recovery.[16] Since the restructuring by intermittent events is a time-dependent process, full elastic recovery is delayed.

How intermittent dynamics in coarsening foams lead to power-law rheology has been recently established in simulations,[30] where both bubble dynamics and stress fluctuations were monitored. In agreement with recent experiments,[31] the mean-square displacement of the bubbles was found to be superdiffusive, while the stress fluctuations were diffusive. Using microrheology arguments,[32,33] the authors show that these two features combined lead to $G^*(\omega) \sim \omega^\alpha$ with $\alpha = 0.17$,[30] in good agreement with our experimental findings. However, the simulation results do not account for the contribution of the final relaxation observed experimentally and instead resemble the response for $\tau_R \to \infty$ shown in Fig. 1(b). A similar response is also predicted in SGR for fully equilibrated systems with a mechanical noise temperature just above the glass transition.[23–25] However, for partially equilibrated systems, SGR predicts a creep behavior similar to that observed experimentally for timescales below the age of the system.[25] Beyond this limit, the transition between equilibrated and nonequilibrated creep modes leads to an age-dependent transient relaxation, followed by a long time power-law creep.[25] For our foam, we find such a transition when performing long experiments during which the bubble size significantly increases such that the quasi-stationary conditions realized in the experiments presented here no longer apply; this work will be presented in a future publication.

Finally, let us note that the frequency dependence of the loss modulus of our foams bears resemblances with the "excess wings" in the dielectric loss of deeply supercooled molecular systems.[34] Indeed, the mechanical analog $\chi''(\omega) = G''(\omega)/G_\infty$ shown in Fig. 3(b) displays a low-frequency relaxation peak with a "wing" at higher frequencies. For supercooled liquids, recent simulations attribute the wing to rare and localized rearrangement events, while the peak relates to the growth by "dynamical facilitation" of regions in the material that have already rearranged.[35,36] Within the framework of foam mechanics, the wing can be considered corresponding to delayed elastic storage, consistent with the idea that the few intermittent events occurring within a short lag time will lead to an increase of strain in a matrix that has not yet been restructured—a contribution that can be fully recovered. Whether the final relaxation of foam can be attributed to a growth of already rearranged zones or to a percolation of the rearranged zones has yet to be established.

Let us close by highlighting the somewhat unique features of foams. They are over-jammed systems that are to be considered athermal due to the mesoscopic size of the bubbles. However, unlike other over-jammed systems, such as compressed emulsions,[37] they structurally relax due to dynamics driven by coarsening. These features confer foams the properties of systems that are near the jamming transition, where mechanical noise effectively leads to local structural relaxation. However, while the properties of jammed materials sensitively depend on shear history,[38,39] the coarsening process in foam naturally erases shear history in time by constantly re-injecting mechanical noise into the system. This makes foams ideal systems to study the mechanics of almost jammed systems in the linear range.[40]

## SUPPLEMENTARY MATERIAL

See the supplementary material for details on foam evolution by coarsening and the linearity of all rheological tests performed.

## ACKNOWLEDGMENTS

We thank Emanuela Del Gado, Jan Vermant, and John Crocker for very helpful discussions. Financial support from the Swiss National Science Foundation (Grant No. 200021_172514) is gratefully acknowledged.

## AUTHOR DECLARATIONS

### Conflict of Interest

The authors have no conflicts to disclose.

## DATA AVAILABILITY

The data that support the findings of this study are openly available in Zenodo at http://doi.org/10.5281/zenodo.6381567.

# Delayed elastic contributions to the viscoelastic response of foams

– Supplementary material –

François A. Lavergne[1*], Peter Sollich[2,3] and Véronique Trappe[1]

[1] Department of Physics, University of Fribourg, Chemin du Musée 3, 1700 Fribourg, Switzerland
[2] Institute for Theoretical Physics, University of Göttingen, Friedrich-Hund-Platz 1, 37077 Göttingen, Germany
[3] Department of Mathematics, King's College London, Strand, London WC2R 2LS, United Kingdom
* francois.lavergne@unifr.ch

**Evolution of bubble characteristics and its impact on foam elasticity**

To be able to assess the impact of foam evolution on our rheological tests, we characterize both the bubble size distribution and the elastic modulus as a function of the foam age $t_w$.

The age-dependent bubble size distribution $f(R)$ is characterised using the method described by Gaillard *et al.* [1], with results shown in Fig. S1 (a). For coarsening foam, we generally expect the system to reach a stage where such distribution is self-similar, while the mean bubble size $\bar{R}$ keeps increasing [2,3]. For our foam, we find that for $t_w \geq 3000$ s, $f(R)$ indeed collapses onto a unique master-curve when renormalizing by the mean bubble size $\bar{R}$, as shown in the inset of Fig. S1(a).

As shown in Fig. S1(b), we find that the age dependence of the mean bubble size is well described by

$$\bar{R}(t_w) = \bar{R}(0)(1 + t_w/t_0)^z, \quad (S1)$$

with $\bar{R}(0) \approx 20$ μm, $t_0 \approx 820$ s and $z \approx 0.49$. This result is consistent with $\bar{R} \sim t_w^{1/2}$ expected for bubble coarsening of dry foams in the scaling regime [2,3], and is in good agreement with findings of previous investigations on Gillette foams [4–8].

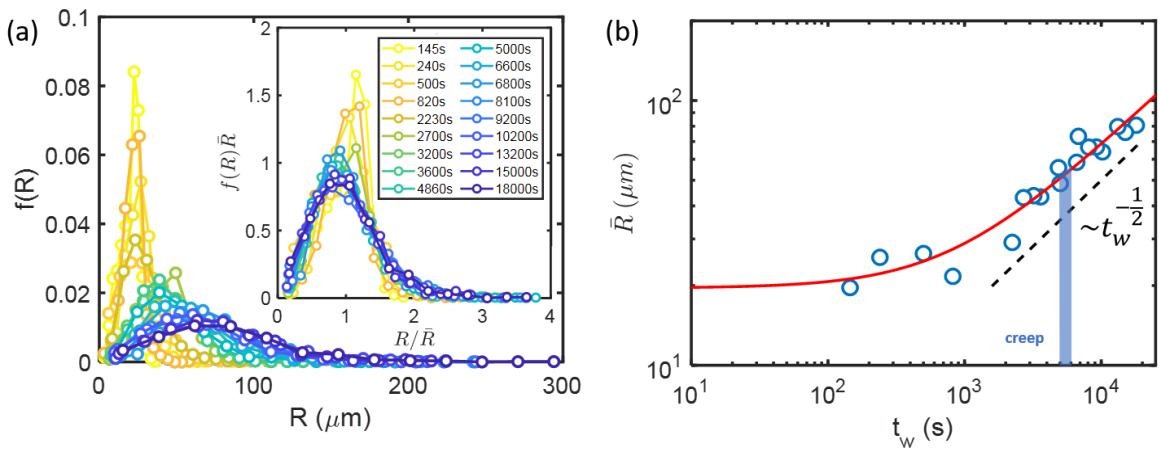

**Figure S1: (a)** Bubble radius distribution for different foam ages (see legend). Inset: rescaled distribution displaying self-similarity for $t_w \geq 3000$ s. **(b)** Mean bubble radius as a function of foam age and fit to Eq. (S1) (red line). For $t_w$ sufficiently large, $\bar{R} \sim t_w^{1/2}$ as expected for bubble coarsening. The blue section indicates the time lap over which creep tests are performed.

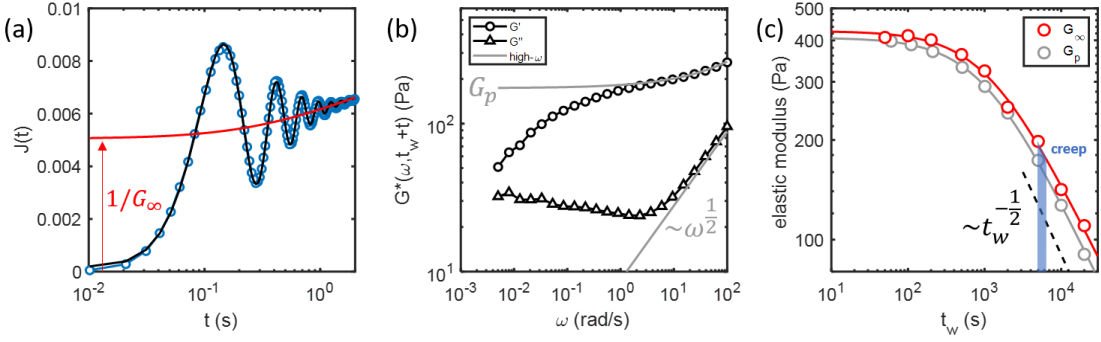

**Figure S2: (a)** Short time behaviour of creep compliance for $t_w = 5000$ s. The black line is the fit of the data to Eq. 1 of the main text, including the expression accounting for ringing in Eq. (S2). The red line corresponds to the fit result corrected for ringing, which is shown in the main text (Fig. 2). **(b)** Frequency-dependence of $G'$ (circles) and $G''$ (triangles) for $t_w = 5000$ s. Data are not corrected for bubble evolution during the experiment. Grey lines correspond to fits to Eq. (S4) in the high frequency regime ($\omega = 10 - 100$ rad/s). **(c)** Age-dependent high frequency modulus $G_\infty$ (red circles) as obtained from fits of the creep compliance (see (a) as an example). Age-dependent apparent plateau modulus $G_p$ (grey circles) as obtained from fits of the high frequency response (see (b) as an example). The red and grey lines are fits to Eq. (S3). For $t_w$ sufficiently large, $G_\infty$ and $G_p \sim t_w^{-1/2}$, as expected for bubble coarsening. The blue section indicates the time lap over which creep tests are performed.

For our experiments starting at $t_w = 5000$ s, these findings show (i) that the foam is well within the scaling regime, and (ii) that the mean bubble size increases from $\bar{R} \approx 52$ μm to $\bar{R} \approx 56$ μm over the duration of a creep experiment ($t_c = 1000$ s), as denoted in Fig. S1(b). This relatively small increase of $\approx 8\%$ ensures quasi-stationary conditions as mentioned in the main text.

As the elastic modulus of a foam typically scales as the inverse of the bubble radius [2,9], we expect the foam elasticity to decrease as a function of the foam age. To assess this dependence, we measure the elastic modulus of the foam at different ages in two different ways, using creep experiments on the one hand, and oscillatory strain experiments on the other hand.

For the creep experiments, we exploit the creep ringing effect observed due to tool inertia. As an example, we show the data obtained for $t_w = 5000$ s in Fig. S2(a). Following reference [10], we fit the short-time oscillations of the compliance to

$$J(t \to 0) \approx \frac{1}{G_\infty}(1 - e^{-At}\sin(wt + p)), \qquad (S2)$$

where $A$, $w$, and $p$ are ringing parameters, and $G_\infty$ is a measure of the high frequency modulus. We repeat this operation at different foam ages and report $G_\infty(t_w)$ in Fig. S2(c). As expected from the evolution of the bubble size, $G_\infty(t_w)$ decreases with $t_w$ and is well described by

$$G_\infty(t_w) = \frac{G_\infty(0)}{(1 + t_w/t_0)^z}, \qquad (S3)$$

with $G_\infty(0) \approx 430$ Pa, $t_0 \approx 1160$ s, and $z \approx 0.48$.

The method based on oscillatory experiments consists of fitting the high frequency-behaviour ($\omega \gtrsim 10$ rad/s) of the complex modulus $G^*(\omega) = G'(\omega) + iG''(\omega)$ to the form proposed in references [7,11]

$$G^*(\omega, t_w) = G_p(t_w)(1 + \sqrt{i\omega/\omega_n(t_w)}), \qquad (S4)$$

where $G_p$ is an apparent plateau modulus and $\omega_n$ a characteristic frequency. We then retrieve the values of $G'$ and $G''$, as shown for $t_w = 5000$ s in Fig. S2(b). Since the acquisition of high frequency data (100 - 10 rad/s) is fast, the foam elasticity barely changes during the acquisition performed by decreasing the frequency from 100 rad/s downwards. The foam age is thus unambiguously defined by $t_w$ at the start of the experiment. As shown in Fig. S2 (c), $G_p(t_w)$ is somewhat smaller than $G_\infty(t_w)$, but exhibits the same age-dependence. Indeed, a fit to the form of Eq. S3 yields $t_0 \approx 1000s$, and $z \approx 0.49$ in close agreement with the fit parameters of $G_\infty(t_w)$, while $G_p(0) \approx 408$ Pa is somewhat smaller than $G_\infty(0)$.

As already mentioned for the evolution of the bubble size, these findings show that the modulus does not change by more than $\approx 8\%$ during our creep experiments ($t_c = 1000$ s), such that we can consider the foam to be in a quasi-stationary state during the test.

However, the acquisition of the frequency dependent moduli down to $\omega = 5 \times 10^{-3}$ rad/s shown in Fig. S2(b) requires a measuring time of 5000 s. During this time laps the foam elasticity decays by $\approx 25\%$. As this is non-negligible, the resulting complex modulus $G^*(\omega, t_w + t)$ depends on the time $t$ passed since the start of the experiment at $t_w$. To correct for this, we exploit the known dependence of $G_p(t_w)$ on the foam age to define a stationary modulus as

$$G^*(\omega, t_w) = G^*(\omega, t_w + t) G_p(t_w)/G_p(t_w + t), \tag{S5}$$

from which we obtain $G'$ and $G''$ reported in Fig. 1(b) of the paper. Note that performing the correction with $G_\infty$ instead of $G_p$ leads to the same result, as both moduli exhibit the same age dependence. To test the robustness of this correction we perform an experiment in which the frequency is increased from low to high frequencies, which is the inverse of our standard protocol. The raw data obtained for both type of experiments, each started at $t_w = 5000$ s, display significant differences, as shown in Fig. S3(a). After correction, both data sets are in good agreement, as shown in Fig. S3(b). This validates our correction method, which permits to retrieve the frequency behaviour of the foam in quasi-stationary conditions.

Let us finally remark that we accounted for creep ringing in our analysis of the creep compliance described in the paper. We however do not include the ringing term in Eq. 1 in order to focus on the characteristics of the foam mechanics, rather than on features set by the tool inertia.

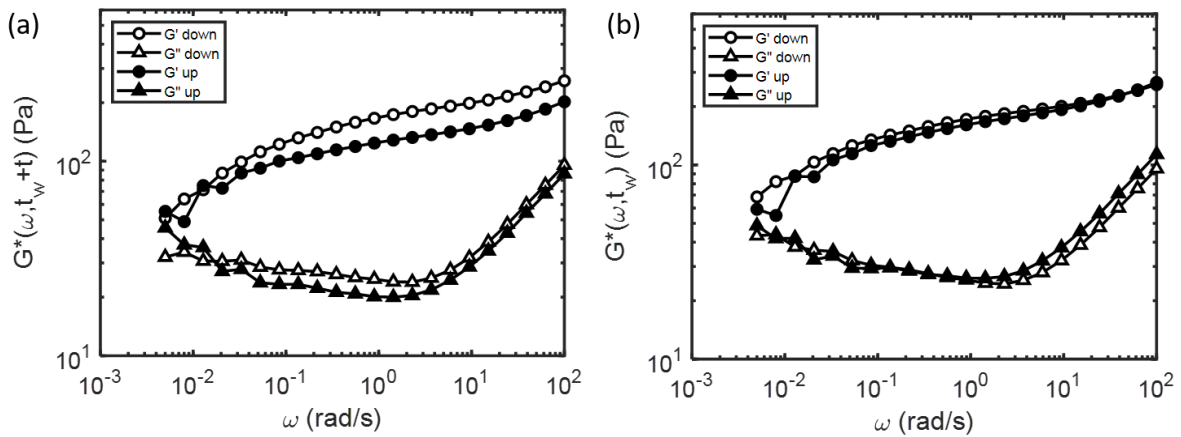

**Figure S3: (a)** Uncorrected frequency sweep data obtained by decreasing the frequency (open symbols) and by increasing the frequency (filled symbols). The foam age at the start of the experiment is $t_w = 5000$ s in both cases. **(b)** Stationary frequency sweep data obtained by correcting the data in (a) using Eq. (S5). The good agreement between both procedures validates the correction scheme used.

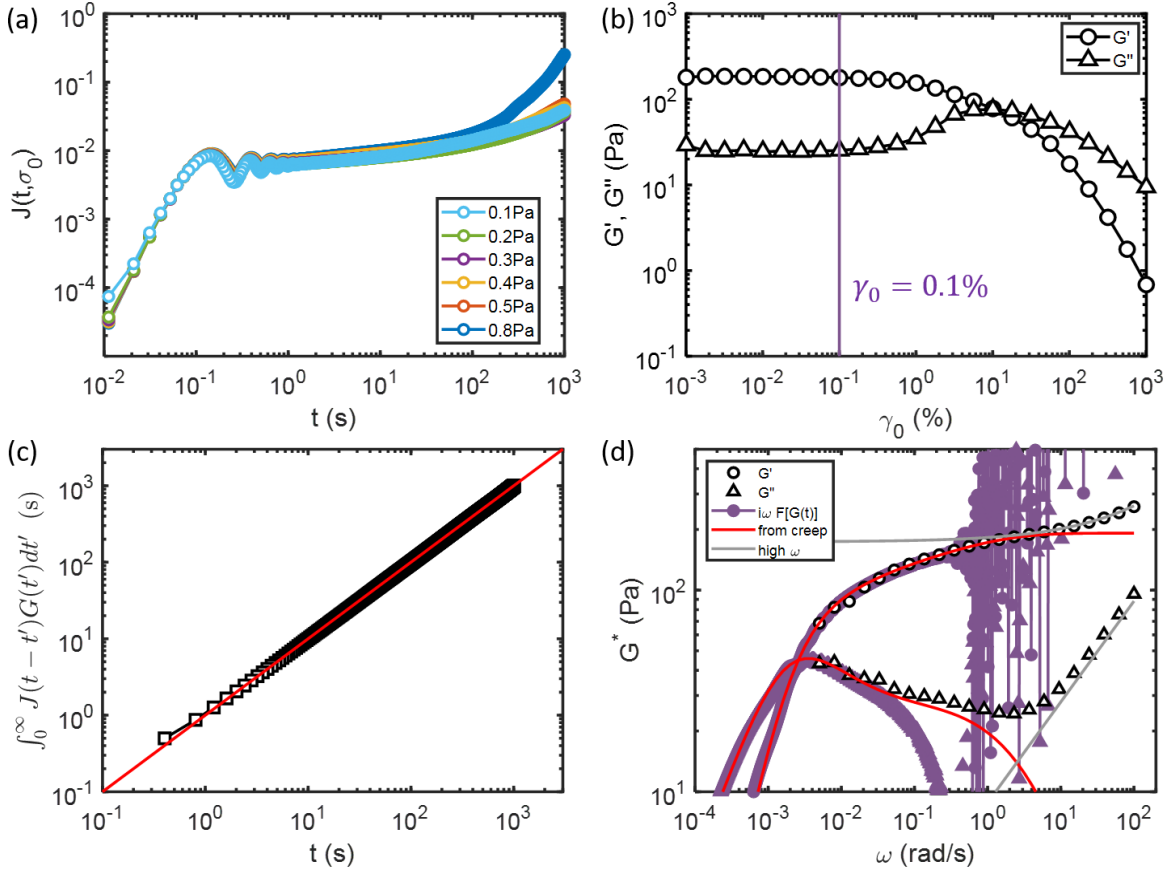

**Figure S4:** All tests in the figure are performed at $t_w = 5000$ s. **(a)** Creep compliance obtained for different applied stresses ($\sigma_0 = 0.1 - 0.8$ Pa, see legend). The limit of linearity is reached in the region $0.5$ Pa $< \sigma_0 < 0.8$ Pa. **(b)** Strain amplitude dependence of $G'$ (circles) and $G''$ (triangles) as measured for $\omega = 5$ rad/s. The limit of linearity is reached at $\gamma_0 \approx 0.3$ %. **(c)** Convolution product of the creep compliance and the relaxation modulus reported in the main text. The excellent agreement with Eq. (S6) (red line) shows that linearity holds at all times during the experiments. **(d)** Stationary frequency sweep data reported in the paper (open symbols) and Fourier transform of the stress relaxation data (purple symbols). The good agreement at lower frequency shows that linearity is fulfilled. The transform fails at high frequency due to the limited time-resolution of the stress relaxation test. The description obtained from creep and discussed in the paper is shown as red lines. Grey lines correspond to the description of the high frequency data according Eq. (S4).

**Ensuring linearity**

To warrant experiments in the linear range, we explore creep at different stresses and ensure that the stress chosen for our creep experiments ($\sigma_0 = 0.2$ Pa) is within the range of stresses where the compliance remains unchanged, see Fig. S4(a). For the choice of the strain used in stress relaxation and oscillatory strain experiments, we measure $G'$ and $G''$ at a fixed frequency of 5 rad/s and vary the amplitude with results shown Fig. S4(b). We chose to work with $\gamma_0 = 0.1\%$, which is a strain within the linear range, while ensuring sufficient torque to obtain reliable data at all frequencies investigated.

A very powerful cross-check of linearity consists in verifying that the following relation holds [12]

$$\int_0^\infty J(t-t')G(t')dt' = t. \qquad (S6)$$

As shown in Fig. S4(c), the convolution integral on the left-hand side computed from the experimental data of $J$ and $G$ closely matches the line $y = t$. This shows unambiguously that both $J$ and $G$ are probed within the linear response regime at all times during the tests.

Finally, the linearity of the frequency sweep can be checked by converting the relaxation modulus into the complex modulus as $G^*(\omega) = i\omega \mathcal{F}[G(t)]$, where $\mathcal{F}[G(t)] = \int_0^\infty G(t')e^{-i\omega t'} dt'$ denotes the one-sided Fourier transform [12,13]. As shown in Fig. S4(d), the data obtained are rather noisy at high frequency due to the limited time resolution of our stress relaxation test. At lower frequencies however, the data obtained by Fourier transform are in good agreement with the frequency sweep data.

Let us finally remark that our approach based on fitting the creep compliance to recast the low-frequency relaxation has many advantages, as shown in Fig. S4(d) and Fig. 1(b) of the main text:

- the results extend to a mid-frequency range ($0.1 - 1$ rad/s) where the simple transform $i\omega\mathcal{F}[G(t)]$ fails due to limited time resolution in relaxation tests.
- the development of $G'$ nicely connects to the high frequency description by Eq. (S4).
- the possibility to analytically treat the description of the creep data enables us to compute the results expected for $\tau_R \to \infty$, which are otherwise not accessible.
- most importantly, while the transform $i\omega\mathcal{F}[G(t)]$ works satisfactorily at low frequency, it does not convey any physical information about the processes involved. By contrast, our approach based on creep and recovery highlights the role of delayed elastic storage at low frequencies.